\documentclass[acmsmall,screen]{acmart}

\usepackage{graphicx}
\usepackage{textcomp}
\usepackage{array}
\usepackage{epstopdf}
\usepackage{subfigure}
\usepackage{caption}
\usepackage{multirow}
\usepackage{titlesec}
\usepackage{setspace}
\usepackage{algorithm}
\usepackage{algorithmic}
\usepackage{listings}
\usepackage{xcolor}
\usepackage[misc]{ifsym}

\usepackage{url}

\usepackage{xcolor}
\newcommand{\mr}[1]{\textcolor{black}{#1}}

\newcommand{\ie}{\textit{i.e.,} }
\newcommand{\eg}{\textit{e.g.,} }

\newcommand\opt[1]{}
\newcommand\find[1]{}

\newcommand{\ls}[1]
   {\dimen0=\fontdimen6\the\font 
    \lineskip=#1\dimen0
    \advance\lineskip.5\fontdimen5\the\font
    \advance\lineskip-\dimen0
    \lineskiplimit=.9\lineskip
    \baselineskip=\lineskip
    \advance\baselineskip\dimen0
    \normallineskip\lineskip
    \normallineskiplimit\lineskiplimit
    \normalbaselineskip\baselineskip
    \ignorespaces
   }

\newenvironment{smalldescription}{
   \setlength{\topsep}{0pt}
   \setlength{\partopsep}{0pt}
   \setlength{\parskip}{0pt}
   \begin{description}
   \setlength{\leftmargin}{.2in}
   \setlength{\parsep}{0pt}
   \setlength{\parskip}{0pt}
   \setlength{\itemsep}{0pt}}{\end{description}}

\titlespacing*{\section} {4pt}{5pt}{4pt}
\titlespacing*{\subsection} {3pt}{3pt}{3pt}
\titlespacing*{\subsubsection} {2pt}{3pt}{2pt}
\setcopyright{rightsretained}
\acmDOI{10.1145/3660777}
\acmYear{2024}
\copyrightyear{2024}
\acmSubmissionID{fse24main-p109-p}
\acmJournal{PACMSE}
\acmVolume{1}
\acmNumber{FSE}
\acmArticle{70}
\acmMonth{7}
\received{2023-09-28}
\received[accepted]{2024-04-16}

\begin{document}

\title{CC2Vec: Combining Typed Tokens with Contrastive Learning for Effective Code Clone Detection}

\author{Shihan Dou}
\orcid{0009-0002-6013-3035}
\affiliation{%
  \institution{Fudan University}
  \city{Shanghai}
  \country{China}
}
\email{shdou21@m.fudan.edu.cn}

\author{Yueming Wu}
\orcid{0000-0002-1515-3558}
\authornote{Yueming Wu is the corresponding author.}
\affiliation{%
  \institution{Nanyang Technological University}
  \city{Singapore}
  \country{Singapore}
}
\email{wuyueming21@gmail.com}

\author{Haoxiang Jia}
\orcid{0009-0005-6027-231X}
\affiliation{%
  \institution{Huazhong University of Science and Technology}
  \city{Wuhan}
  \country{China}
}
\email{haoxiangjia@hust.edu.cn}

\author{Yuhao Zhou}
\orcid{0009-0008-8665-3999}
\affiliation{%
  \institution{Fudan University}
  \city{Shanghai}
  \country{China}
}
\email{zhouyh21@m.fudan.edu.cn}

\author{Yan Liu}
\orcid{0009-0002-5741-336X}
\affiliation{%
  \institution{Fudan University}
  \city{Shanghai}
  \country{China}
}
\email{yliu22@m.fudan.edu.cn}

\author{Yang Liu}
\orcid{0000-0001-7300-9215}
\affiliation{%
  \institution{Nanyang Technological University}
  \city{Singapore}
  \country{Singapore}
}
\email{yangliu@ntu.edu.sg}



\begin{abstract}
With the development of the open source community, the code is often copied, spread, and evolved in multiple software systems, which brings uncertainty and risk to the software system (e.g., bug propagation and copyright infringement). Therefore, it is important to conduct code clone detection to discover similar code pairs. Many approaches have been proposed to detect code clones where token-based tools can scale to big code. However, due to the lack of program details, they cannot handle more complicated code clones, \ie semantic code clones. In this paper, we introduce \emph{CC2Vec}, a novel code encoding method designed to swiftly identify simple code clones while also enhancing the capability for semantic code clone detection. To retain the program details between tokens, \emph{CC2Vec} divides them into different categories (\ie typed tokens) according to the syntactic types and then applies two self-attention mechanism layers to encode them. To resist changes in the code structure of semantic code clones, \emph{CC2Vec} performs contrastive learning to reduce the differences introduced by different code implementations. We evaluate \emph{CC2Vec} on two widely used datasets (\ie BigCloneBench and Google Code Jam) and the results report that our method can effectively detect simple code clones. In addition, \emph{CC2Vec} not only attains comparable performance to widely used semantic code clone detection systems such as \emph{ASTNN}, \emph{SCDetector}, and \emph{FCCA} by simply fine-tuning, but also significantly surpasses these methods in both detection efficiency.

\end{abstract}

\maketitle

\section{Introduction}
\mr{
Code clone, also known as duplicate code or similar code, refers to two or more identical or similar source code fragments that exist in the code base.
According to the syntactic or semantic level differences, code clones can be classified into four types \cite{roy2007type1_4, bellon2007type1_4}.
Type-1 to Type-3 code clones belong to syntactic code clones, while Type-4 code clones are semantic code clones.
Most of the existing code clone detection methods aim to detect syntactic code clones because these clones are easier to identify compared to semantic code clones.
}
For example, to detect the first two types of code clones, \emph{CCFinder} \cite{kamiya2002ccfinder} first performs lexical analysis to extract the token sequence of a code snippet.
After obtaining all tokens, it conducts several code transformations to convert the tokens into corresponding forms and then uses the transformed tokens to finish code clone detection.
Since \emph{CCFinder} \cite{kamiya2002ccfinder} only applies certain simple code transformations, it cannot detect Type-3 code clones.
To address the issue, other token-based methods are designed by analyzing the token features between the two codes within a pair from different perspectives.
For example, \emph{SourcererCC} \cite{sajnani2016sourcerercc} computes the overlap ratio of tokens to detect near-miss Type-3 code clones.
However, due to the lack of preservation of program semantics, these token-based methods do badly in detecting semantic code clones since the code structure may change a lot after using another way of implementation.
For example, when implementing the same loop-related functionality, some developers may use \emph{for} statements while others may prefer to use \emph{while} statements.

To tackle the issue, researchers perform complex program analysis to transform the code fragments into different intermediate representations such as \emph{abstract syntax tree} (AST) \cite{zhang2019astnn} and \emph{control flow graph} (CFG) \cite{wu2020scdetector}.
These intermediate representations can maintain the program semantics from different perspectives.
Empirical experiments have demonstrated the capability of tree-based and graph-based methods \cite{krinke2001duplix, komondoor2001pdgdup, wang2017ccsharp} on detecting semantic code clones.
However, both tree processing and graph analysis are time-consuming, making it difficult for them to find semantic code clones efficiently.
A recent report \cite{opensource} has shown that open source has become an unstoppable trend, leading to an increasing scale of code reuse.
These cloned codes can bring uncertainty and risks to the software system, such as bug propagation, and copyright infringement \cite{opensource1}.
Therefore, there is an urgent need to develop a tool that can be used for large-scale semantic clone detection.

\mr{
In this paper, we propose \emph{CC2Vec}, a novel code encoding approach to efficiently detect syntactic code clones, while enhancing the ability of semantic code clone detection by further combining with a few neural networks.
}
Specifically, we mainly address two challenges.
\begin{itemize}
    \item \emph{Challenge 1: How to retain the program semantics by purely analyzing the source code tokens?}
    \item \emph{Challenge 2: \mr{How to differentiate codes based on their functionalities instead of their structures?}}
\end{itemize}

To address the first challenge, we parse the source code into corresponding tokens and divide them into 15 categories based on syntactic types by lexical analysis.
We call tokens in different categories as \emph{typed tokens} and observe that these typed tokens have different weights in detecting code clones (Details are in Section~\ref{sec:background}).
Based on the observation, we first apply one self-attention mechanism layer to encode the tokens in the same category into a vector representation.
By this, the source code is converted into 15 vectors, each corresponding to a token category.
After obtaining 15 vectors, we then apply another self-attention mechanism layer to encode them into one vector, which is the output of this phase. 
Using self-attention mechanism layers can not only interpret the detection results but also preserve the potential relationships between tokens.
These relationships may represent some program details of a method.

To solve the second challenge, we conduct contrastive learning to train the program encoder.
\mr{
The use of contrastive learning is to maximize the similarity between representations of a code snippet and its clone codes (\ie positive samples), while minimizing the representations' similarity between this code snippet and its non-clone codes (\ie negative samples).
}
As for semantic code clones, although their code structures may change a lot, the implemented functionality remains unchanged.
In other words, they can be treated as clone samples of each other.
Therefore, we can use contrastive learning to decrease the distance between semantic code pairs while enlarging the difference between dissimilar code pairs.
After contrastive learning, the learned encoder \emph{CC2Vec} is robust to code changes introduced by semantic code cloning, making it possible to distinguish semantic code clones although they are syntactically different.

\mr{
We design a token-based code clone detector by simply calculating the cosine similarity of the two codes' vector encoded by \emph{CC2Vec}, to detect syntactic code clones efficiently.
In addition, we can effectively detect more complicated code clones (\ie semantic code clones) by combining \emph{CC2Vec} with a few neural networks.
}
To examine the ability of \emph{CC2Vec}, we conduct evaluations on two widely used datasets, namely BigCloneBench \cite{big, svajlenko2014big} and Google Code Jam \cite{gcj}.
\mr{
Compared to five pretrain-based methods, \emph{CC2Vec} can effectively capture the tokens relationship information by pre-training from the training code corpus to achieve an improvement in detecting code clones.
}
Compare to nine traditional code clone detectors, the recall of \emph{CC2Vec} is 64\% while the comparative tools can only maintain very low recall, respectively.
Compared to six deep-learning-based code clone detectors, \emph{CC2Vec} can achieve the best F1 score when using only a simple three-layer neural network as the classifier.
As for scalability, \emph{CC2Vec} outperforms most pretrain-based and deep learning-based methods in terms of runtime efficiency.
\emph{CC2Vec} is about 100 times faster than \emph{ASTNN} in predicting code clone pairs \footnote{\ Our tool is available on our website: \emph{\href{https://github.com/CC2Vector/CC2Vec}{https://github.com/CC2Vector/CC2Vec}}.}.

\par In summary, this paper makes the following contributions:
\mr{
\begin{itemize}  
  \item We introduce a novel encoding approach \emph{CC2Vec} to enhance the ability on code clone detection by using contrastive learning.
  \emph{CC2Vec} applies two self-attention mechanisms layers to encode source code tokens and contrastive learning to learn a robust encoder.
  \item We improve the detection accuracy of syntactic code clones by simply computing the cosine similarity of two codes' vectors encoding by \emph{CC2Vec}.
  Additionally, we can also effectively detect the semantic code clones by combining \emph{CC2Vec} with a few neural networks.  
  \item We check the ability of \emph{CC2Vec} by conducting experiments on two widely used datasets (\ie BigCloneBench and Google Code Jam).
  Experimental results show that our proposed detectors based on \emph{CC2Vec} can achieve comparable performance to widely used code clone detectors with efficient training and detecting phase.
\end{itemize}
}


\par \noindent \textbf{Paper organization.} The remainder of the paper is organized as follows. \mr{Section 2 presents our background.} Section 3 introduces our system. Section 4 reports the experimental results. Section 5 shows our discussions. Section 6 describes the related work. Section 7 concludes the present paper.

\section{Background}
\label{sec:background}

\par
\mr{
In this section, we first discuss four types of code clones and then clarify the key insight of \emph{CC2Vec}.
}

\subsection{\mr{Clone Types}}

As aforementioned, code clones can be divided into four types \cite{roy2007type1_4, bellon2007type1_4}. The first three types belong to syntactic code clone while the last type is semantic code clone.
\begin{itemize}
  \item \textbf{Type-1}: 
  The same code snippets, except for differences in white space and comments.
  \item \textbf{Type-2}: 
  The same code snippets, except for differences in identifier names, literal values, white space, and comments.
  \item \textbf{Type-3}: 
  Syntactically similar code snippets that differ at the statement level. 
  The snippets have statements added, modified and/or removed.
  \item \textbf{Type-4}: 
  Syntactically dissimilar code snippets that implement the same functionality.
\end{itemize}

\begin{figure}[htbp]
\centerline{\includegraphics[width=0.98\textwidth]{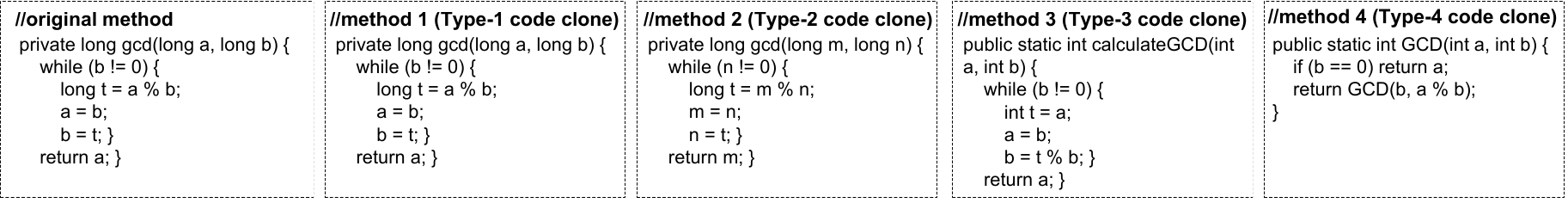}}
\caption{Examples of four code clone types}
\label{fig:clonetype}
\end{figure}

To illustrate the differences between four types of code clones, we present one method and its four types of code clones in Figure \ref{fig:clonetype}.
They implement the same functionality which is to calculate the greatest common divisor of two numbers.
Method 1 is the Type-1 code clone of the original method, with no changes between their source code.
Method 2 is the Type-2 code clone of the original method, and the difference between them is only in the identifiers name (\ie $m$ and $n$ instead of $a$ and $b$). 
The few code changes make them easy to detect.
Method 3 is the Type-3 code clone of the original method, it differs at the statement level.
Their method names and parameter types are different and the order of the statements also changes a little.
This type of code clone is more difficult to detect than the previous two types.
Method 4 is the Type-4 code clone of the original method, it implements the same functionality in a different way.
This type of clone is also called semantic clone and is the most difficult to discover since the code structure may change a lot.

\subsection{Motivation}

\begin{figure}[htbp]
\centerline{\includegraphics[width=0.85\textwidth]{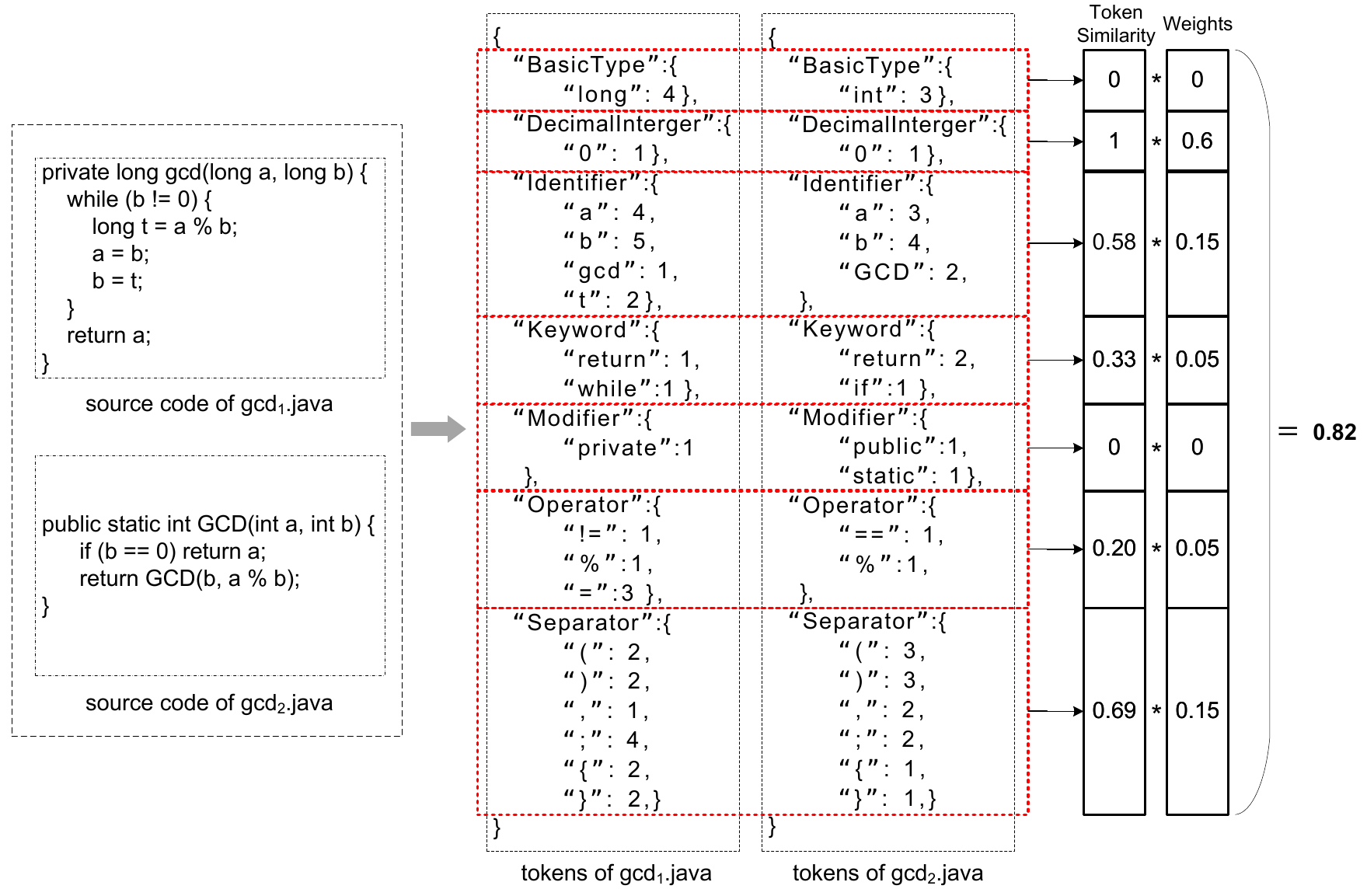}}
\caption{
\mr{Source code and corresponding tokens of \emph{gcd$_1$.java} and \emph{gcd$_2$.java} after lexical analysis}
}
\label{fig:motivation}
\vspace{-1em}
\end{figure}

We propose an example to illustrate the key insight of our method.
This example is a clone pair in BigCloneBench \cite{big}, it consists of two methods and their functionalities are both to compute the greatest common divisor of two integers.
As shown in Figure \ref{fig:motivation}, although they are syntactically dissimilar, the functionalities are the same.
Therefore, they can be treated as a semantic clone pair.

As aforementioned, \emph{SourcererCC} \cite{sajnani2016sourcerercc} is one of the most scalable token-based code clone detector, which can scale to analyze more than 428 million files on Github \cite{lopes2017dejavu}.
\mr{It uses a simple overlap similarity calculation to measure the distance between two methods.}
For example, given two methods \emph{M$_1$} and \emph{M$_2$}, the overlapping similarity $S(M_1, M_2)$ is computed as the quotient of the number of same tokens shared by \emph{M$_1$} and \emph{M$_2$} and the maximum number of tokens in \emph{M$_1$} and \emph{M$_2$}, which can be written as $S(M_1, M_2) = \frac{|M_1 \bigcap M_2|}{max(|M_1|, |M_2|)}$.


To compute the overlapping similarity of methods in List 1 and List 2, we first perform lexical analysis to collect the corresponding source code tokens.
Then the number of the same tokens shared by these two methods is obtained by statistical analysis.
After completing the analysis, we find that there are 19 tokens shared by \emph{gcd$_1$.java} and \emph{gcd$_2$.java}.
Therefore, the similarity computed by \emph{SourcererCC} is 0.5 (\ie 19/38=0.5) since the maximum number of tokens of \emph{gcd$_1$.java} and \emph{gcd$_2$.java} is 38.
Because 0.5 is smaller than \mr{0.7 which is the default similarity threshold in \emph{SourcererCC}}, \emph{gcd$_1$.java} and \emph{gcd$_2$.java} will not be reported as a clone pair.
In other words, it will be a false negative when we use \emph{SourcererCC} to detect \emph{gcd$_1$.java} and \emph{gcd$_2$.java}.

To detect the semantic code pair, we conduct further analysis on their source code tokens.
\mr{
We observe that these tokens can be divided into different categories.
Figure \ref{fig:motivation} presents the tokens in different categories of \emph{gcd$_1$.java} and \emph{gcd$_2$.java} after applying lexical analysis.
Through the results, we can see that the tokens in some categories are similar, while the tokens in certain categories are different.
}
For example, only four ``\emph{long}'' tokens belong to \emph{BasicType} category in \emph{gcd$_1$.java}, while the tokens belonging to \emph{BasicType} category in \emph{gcd$_2$.java} are three ``\emph{int}'' tokens.
As for tokens in \emph{DecimalInterger} category, \emph{gcd$_1$.java} and \emph{gcd$_2$.java} are the same, that is, only one ``\emph{0}'' token.
\mr{
To measure the similarity of tokens in different categories, we apply Eq.~\ref{equ:sou} to the tokens within the single category to compute the category-level overlapping similarity of these two methods.
To measure the similarity of tokens in different categories, we compute the above category-level overlapping similarity of these two methods for each token category.
For instance, the overlapping similarity for the \emph{Separator} category can be calculated as 0.69 (\ie (2+2+1+2+1+1)/max((2+2+1+4+2+2), (3+3+2+2+1+1)) = 9/13 = 0.69).
}
After computing all similarities, we find that they are different in different categories.
Some categories have high similarities, while others have low similarities.
In other words, if we assign a higher weight to a category with high similarity and a smaller weight to a category with low similarity, then \emph{gcd$_1$.java} and \emph{gcd$_2$.java} are likely to be detected as a clone pair.
For example, if the weights of seven categories in Figure \ref{fig:motivation} are 0, 0.6, 0.15, 0.05, 0, 0.05, and 0.15, respectively.
\mr{The similarity will be calculated as 0.82 (\ie 0$*$0 + 1$*$0.6 + 0.58$*$0.15 + 0.33$*$0.05 + 0$*$0 + 0.20$*$0.05 + 0.69$*$0.15 = 0.82. In each multiplication, the first value represents the category overlap similarity, and the second value denotes its assigned weight) which is higher than the default similarity threshold (\ie 0.7) in \emph{SourcererCC}.}

\mr{
Therefore, inspired by the observation, we introduce the attention mechanism to design a novel technique to train an encoder that can assign suitable weights for different categories, to detect all kinds of code clones.
The attention mechanism can automatically focus on important tokens, which has the potential to filter the useless tokens and categories in detecting code clones, while concentrating the attention on these useful tokens and categories.
}

\begin{figure*}[htbp]
\centering
\includegraphics[width=0.85\textwidth]{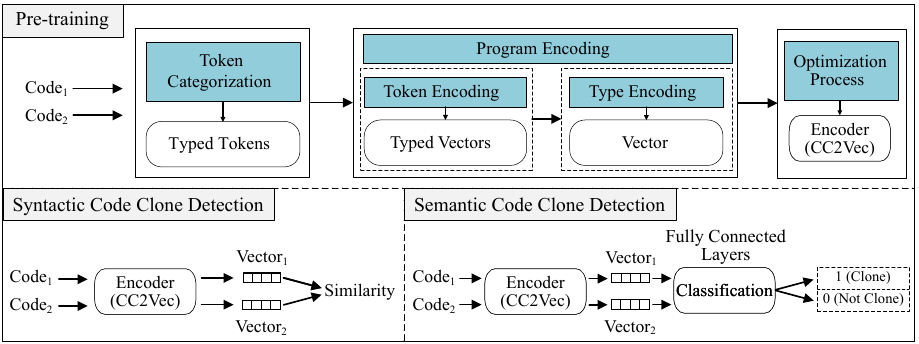}
\caption{\mr{System of \emph{CC2Vec}}}
\label{fig:system}
\end{figure*}

\section{System}

\par
\mr{
In this section, we introduce our code encoding approach, namely \emph{CC2Vec}, and how to utilize it to detect syntactic and semantic code clones.
}

\subsection{\emph{CC2Vec} Overview}
\par As shown in Figure ~\ref{fig:system}, \emph{CC2Vec} is made up of three main components: \emph{Token Categorization}, \emph{Program Encoding}, and \mr{\emph{Optimization Process}}.

\begin{itemize}
  \item 
  \textbf{\emph{Token Categorization}}: 
  Given the source code of a method, we first apply lexical analysis to parse them into tokens with corresponding categories, namely \emph{typed tokens}.
  The input of this component is a method, while the output is certain typed tokens.
  \item 
  \textbf{\emph{Program Encoding}}: 
  This component consists of two attention mechanism layers. The first layer encodes tokens in each category into a vector, and the second layer encodes all categories into the final vector.
  The input of this component is certain typed tokens, while the output is a vector representation.
  \item \mr{
  \textbf{\emph{Optimization Process}}:
  In this phase, we employ contrastive learning as the optimization approach to train the program encoder which is robust to code changes introduced by different implementations.
  }
\end{itemize}

\begin{table}[htbp]
\caption{Selected 15 token types}
\small
\label{tab:types}
\begin{spacing}{0.9}
\setlength{\tabcolsep}{3mm}{
\begin{tabular}{c}
\toprule
\toprule  
``\emph{Annotation}'', ``\emph{BasicType}'', ``\emph{BinaryInteger}'', ``\emph{Boolean}'', ``\emph{DecimalFloatingPoint}'', ``\emph{Modifier}''\\, ``\emph{Operator}'',
``\emph{DecimalInteger}'', ``\emph{HexFloatingPoint}'', ``\emph{HexInteger}'', ``\emph{Identifier}'',\\ ``\emph{Keyword}'', ``\emph{OctalInteger}'', ``\emph{Separator}'', ``\emph{Null}''\\
\bottomrule
\bottomrule
\end{tabular} }%
\end{spacing}
\end{table}

\subsection{Token Categorization}

This component aims to parse the source code of a method into tokens with corresponding categories (\ie \emph{typed tokens}).
To complete the purpose, we conduct a lexical analysis to analyze the source code to divide it into different categories based on syntactical types.
Since the experimental dataset is BigCloneBench \cite{big} and the programming language is \emph{Java}, we leverage a python library \emph{javalang} \cite{javalang} to accomplish the lexical analysis of \emph{CC2Vec}.
In practice, token categorization aims to divide a method into \emph{typed tokens}, which can be achieved by lexical analysis.
Therefore, \emph{CC2Vec} is not limited to programming languages (\eg \emph{C} and \emph{Java}) because different languages have related lexical analysis tools (\eg \emph{pycparser} \cite{pycparser} for \emph{C} language and \emph{javalang} \cite{javalang} for \emph{Java} language).

\mr{In practice, listing all token types parsed by \emph{javalang} can be challenging due to the potential existence of some types that are used infrequently.}
To identify the most relevant token types, we conducted a lexical analysis on the top 10,000 Java projects on GitHub, ranked by their criticality score. 
This score can be used to describe the impact and importance of an open source project \cite{criticalityscore}.
After our statistical analysis, we found that 14 types collectively constitute over 99.5\% of all tokens. 
Therefore, we choose these 14 types as the final token types and add a \emph{Null} type to represent other types as shown in Table \ref{tab:types}.

\begin{figure}[h]
\centering
\includegraphics[width=0.85\textwidth]{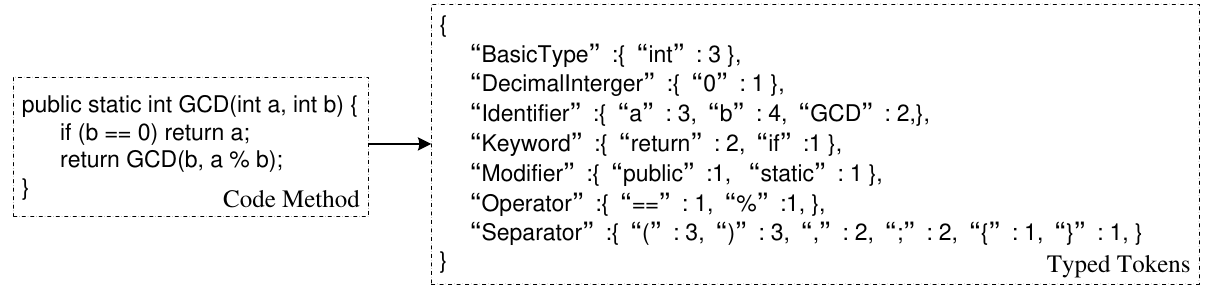}
\caption{Token categorization of \emph{CC2Vec}}
\label{fig:typetokens}
\end{figure}

To better describe the different steps of our system, we present one simple example in Figure \ref{fig:typetokens}.
It shows that tokens of the method in List 2 can be divided into seven categories after applying lexical analysis and each category contains the corresponding tokens and the number of tokens.
For example, in ``\emph{BasicType}'' category, it only contains one token ``\emph{int}'', and the number of tokens ``\emph{int}'' is three.
The output of this component is tokens with corresponding categories, namely \emph{typed tokens}.
\mr{
Since different tokens may determine whether two pieces of code are clones, as discussed in Section~\ref{sec:background}, if we can identify the optimal tokens and increase their weights automatically, we will be able to detect code clones more effectively.
}

\begin{figure}[h]
    \centering
    \begin{minipage}{0.522\textwidth}
        \centering
        \includegraphics[width=\textwidth]{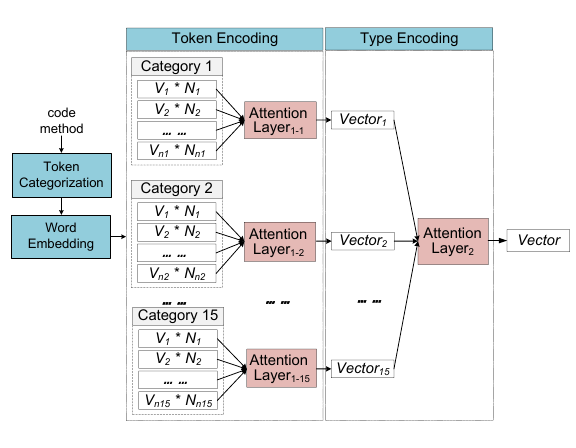}
        \captionof{figure}{Program encoding of \emph{CC2Vec}}
        \label{fig:encoder}
    \end{minipage}%
    \begin{minipage}{0.448\textwidth}
        \centering
        \includegraphics[width=\textwidth]{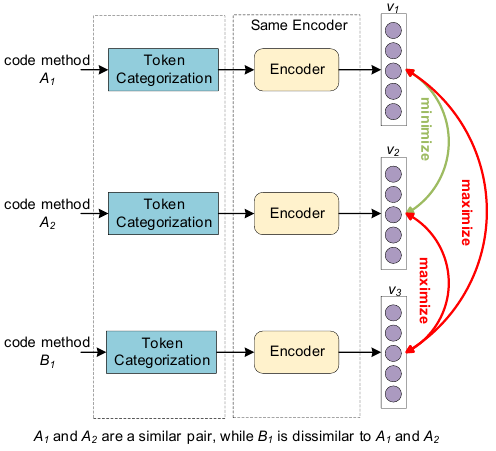}
        \captionof{figure}{Contrastive learning of \emph{CC2Vec}}
        \label{fig:learning}
    \end{minipage}
\end{figure}



\subsection{Program Encoding}

This component aims to encode \emph{typed tokens} into a vector representation.
As shown in Figure \ref{fig:encoder}, it consists of two steps: \emph{Token Encoding} and \emph{Type Encoding}.

\subsubsection{Token Encoding.}
This step aims to encode tokens within a single category into vector representations. To achieve this, we initially employ a word embedding technique to map tokens to their corresponding vector representations. Subsequently, these vectors are input into an attention mechanism layer for analysis. Specifically, to encode tokens with similar meanings into high-dimensional feature spaces with comparable distances, we utilize the source code from BigCloneBench \cite{big} as the training set to train a \emph{Word2Vec} model \cite{word2vec}. \emph{Word2Vec} stands as one of the most widely used techniques for learning word embeddings through shallow neural networks.
After obtaining corresponding vector representations of all tokens in one category, we multiply them by the number of corresponding tokens one by one. 
These multiplied vectors will be fed into one attention mechanism layer to be encoded.
Note that each category is assigned an attention mechanism layer, and the layers are independent of each other.
In other words, we design 15 independent attention mechanism layers to encode tokens in 15 categories, respectively.
Some semantic relationships exist between statements in a method (\eg, data flow), resulting in potential relationships between tokens. 
To consider these potential relationships while program encoding, we use the self-attention mechanism as our encoding layer since it can not only describe the importance or weights but also consider the relationships between tokens. 
Our self-attention mechanism layer consists of one multi-head attention layer and one feed-forward layer.
More details of them are shown in \emph{Transformer} \cite{vaswani2017attention}.
After token encoding, each category can output a vector.

\subsubsection{Type Encoding.}
This step aims to encode vectors obtained from \emph{Token Encoding} step into a vector representation.
Due to some potential relationships between tokens, there are some relationships between categories.
Therefore, we also apply one self-attention mechanism layer to finish the final encoding.
It is worth noting that some methods might not encompass all 15 categories.
For example, after lexical analysis, the method in Figure \ref{fig:typetokens} can only be divided into seven categories.
To ensure appropriate weight allocation across all categories, we assign a value of 0 to vectors corresponding to absent categories.
\mr{
Setting each position of absent types of token to zero, instead of category 0 is intended to ensure that the results of subsequent matrix operations in attention are 0, preventing this token from interfering with others.
}
By this, the inputs of \emph{Type Encoding} step are 15 vectors corresponding to 15 categories.
The self-attention mechanism layer used in this step is the same as that used in \emph{Token Encoding} step.
After \emph{Type Encoding}, we can obtain the final vector representation of a given method.

\subsection{\mr{Optimization Process}}
\label{sec:clone-detection}

Given a dollar bill, it is difficult for us to draw it exactly the same, although we have seen it many times. 
However, we can easily use our drawing to convey the essence of the dollar bill to others \cite{blog_contrastive}.
Drawing inspiration from such commonplace observations, researchers have designed a learning algorithm. Instead of focusing on every sample detail, it extracts enough abstract features to distinguish it from other samples.
This learning algorithm is also known as contrastive learning, its goal is to make the distance between different types of inputs get larger and larger, and the same types of inputs get closer and closer.
Since contrastive learning only needs to learn to distinguish the sample in the feature space of the abstract semantic level, the model and its optimization become simpler and the generalization ability is stronger.
Formally, contrastive learning aims to allow the encoder to generate more unique data features without losing its essence.
For any data sample $x$ and its feature $f(x)$, it intends to complete a task as equation (\ref{equ:con_1}).
\begin{equation}
\small
    \text{SIM}(f(x), f(x^+)) >> \text{SIM}(f(x), f(x^-))
    \label{equ:con_1}
\end{equation}
\mr{Here $x^+$ is a data sample similar to $x$ (referred to as the positive sample) and $x^-$ is a data sample dissimilar to $x$ (referred to as the negative sample).} 
$f(x^+)$, $f(x^-)$ is the feature of $x^+$, $x^-$ separately, and $\text{SIM}(\cdot)$ is a metric to measure the similarity between two features.

\mr{We utilize contrastive learning as the optimization method to train the program encoder.}
Specifically, the model training and optimization process is shown in Figure \ref{fig:learning}, where code method $A_{1}$ and code method $A_{2}$ are a clone pair, and code method $B_{1}$ is dissimilar to code methods $A_{1}$ and $A_{2}$.
\mr
{Therefore, the optimization goal of the model is to reduce the distance between the encoded vectors of the code method and its similar pairs (\ie positive samples), such as $A_{1}$ and $A_{2}$, and increase the distance between the encoded vectors of the code method and its dissimilar pairs (\ie negative samples), such as $A_{1}$ and $B_{1}$.}
As for semantic code clones, although their code structures may change a lot, the implemented functionality remains unchanged.
Therefore, they can be treated as positive samples of each other.

\subsection{\mr{Code Clone Detection}}

\mr{
After completing the training phase by using contrastive learning, the learned encoder can be robust to code changes introduced by different implementations, making it possible to identify the subtle difference between the two code snippets.
The code encoder can be directly used for discovering the simple syntactic code clones, while can detect semantic code clones by continuing fine-tuning. 
}

\mr{
Firstly, \emph{CC2Vec} introduces two self-attention mechanisms layers to capture potential relationships between tokens, which means that it can encode more important information.
Therefore, we can directly design a simple code clone detector to identify code clones efficiently, and easily to scale to big code.
Specifically, for the Type-1 to Type-3 code clones (\ie syntactic code clones), we obtain the two code representations by using \emph{CC2Vec} and calculate the cosine similarity between these two vectors.
If their cosine similarity is higher than 70\%~\cite{sajnani2016sourcerercc}, they will be reported as a clone pair. Otherwise, they will be treated as a non-clone pair.
}

\mr{
Secondly, for the Type-4 code clones (\ie semantic code clones), we combine \emph{CC2Vec} with a few neural networks (\ie fully connected layer) to enhance the ability to detect more complicated samples.
As \emph{CC2Vec} has learned potential information from code data, fine-tuning it with one to three fully connected layers can significantly improve the detection ability on semantic code clones.
Specifically, we obtain the vectors corresponding to each of the two code snippets within a pair and then concatenate them to form a single vector representation.
This vector is then input to multiple fully connected layers to detect whether the code pair is a clone pair or not.
For instance, \emph{CC2Vec-3L} denotes that we connect our pre-trained encoder \emph{CC2Vec} with three fully connected layers.
Notably, we insert a ReLu activation layer between each fully connected layer to ensure the non-linear capability of the neural networks. 
}



\section{Experiments}

\mr{
In this section, we undertake comprehensive experiments to assess \emph{CC2Vec}'s ability to detect code clones from three key perspectives: effectiveness, interpretability, and scalability.
For effectiveness, we not only evaluate \emph{CC2Vec}'s performance in detecting both syntactic and semantic code clones but also demonstrate the contributions of its different components (\ie self-attention and contrastive learning) to code clone detection.
For interpretability, we provide examples to illustrate how \emph{CC2Vec} interprets the detection results, shedding light on its decision-making process.
For scalability, we present a detailed experiment to show the runtime overhead of \emph{CC2Vec}, providing insights into its computational efficiency.
Specifically, we aim to answer the following research questions:
}

\begin{itemize}
  \item \mr{\emph{RQ1: Can CC2Vec detect code clones effectively?}}
  \item \emph{RQ2: How do self-attention mechanism layers and contrastive learning contribute to CC2Vec?}
  \item \emph{RQ3: Can CC2Vec interpret the detection results?}
  \item \emph{RQ4: Can CC2Vec scale to big code?}
\end{itemize}

\subsection{Experimental Settings}

\subsubsection{Dataset.}

To check the ability of \emph{CC2Vec} to detect code clones, we choose BigCloneBench \cite{big} as our first dataset since it is the largest dataset and widely used by many researchers \cite{sajnani2016sourcerercc, zhang2019astnn, wang2018ccaligner}.
Experts assign all pairs in BigCloneBench a clone type according to a similarity score calculated by line-level and token-level after code normalization.
As for Type-3 and Type-4 code clones, they are divided into four parts due to the ambiguous boundary between them, that is, 1) \emph{Very Strongly Type-3} (VST3) with a similarity between [0.9, 1.0), 2) \emph{Strongly Type-3} (ST3) with a similarity between [0.7, 0.9), 3) \emph{Moderately Type-3} (MT3) with a similarity between [0.5, 0.7), and 4) \emph{Weakly Type-3/Type-4} (WT3/T4) with a similarity between [0.0, 0.5).
\mr{
BigCloneBench contains about 270,000 non-clone pairs and over eight million tagged clone pairs. 
Finally, we obtain 48,116 T1 clones, 4,234 T2 clones, 4,577 VST3 clones, 16,818 ST3 clones, 86,341 MT3 clones and 7,943,729 WT3/T4 clones.
To balance our dataset, we select almost 270, 000 clone pairs from them.
Specifically, we select all of the syntactic pairs to balance the number of each code clone pair type, and randomly select 109,914 clone pairs from a total of about eight million semantic clones. 
As a result, there are 48,116 T1 clones, 4,234 T2 clones, 4,577 VST3 clones, 16,818 ST3 clones, 86,341 MT3 clones, and 109,914 WT3/T4 clones in our dataset.
}

\mr{
Similar to previous work \cite{wu2020scdetector}, for the more challenging semantic clones, we also evaluate our proposed method on another widely used dataset namely Google Code Jam \cite{zhao2018deepsim}, which is primarily composed of semantic clones, to better demonstrate the effectiveness of our method.
}
Google Code Jam is derived from an online programming competition held by Google and contains 1,669 projects from 12 different competition problems which are written by different programmers.
So projects of the same competing problem are almost syntactically different but semantically similar, and we treat them as clone pairs. 
Projects that solve different problems are not similar, and we regard them as non-clone pairs.
\mr{
Finally, we obtain 275,570 semantic clone pairs and 1,116,376 non-clone pairs. 
We randomly select 270,000 pairs from all non-clone pairs to balance our dataset.
}

\subsubsection{Implementations.}

For detecting the syntactic code clones, we only need to train a code encoder \emph{CC2Vec}.
\mr{
We use \emph{Word2Vec} \cite{word2vec} to embed the token to the vector.
The size of the embedding layer and the hidden size of the two attention layers are 100.
We average the hidden cell of the attention layer to obtain the output vector, so the vector size is also 100.
During the pre-trained phase, we freeze the embedding layer and only the attention layer is trainable.
}
For contrastive learning, we apply the SupCon loss function \cite{khosla2020supervised} as the training objective.
The temperature in this loss function is set to 0.07.
During the training phase, we use the RMSProp optimization algorithm \cite{RMSProp}, with a momentum of 0.9 and a weight decay of 0.0001, to train the encoder.
The learning rate and the training epochs are set to 0.0001 and 10, respectively.
The training batch size is 64.
After training, we can obtain a program encoder to extract high-level potential features from code snippets. 
The threshold is set to 0.7, based on the previous research~\cite{sajnani2016sourcerercc}, indicating that if the cosine similarity between the two code snippets within a pair exceeds 70\%, they are classified as a clone pair. Otherwise, they are considered a non-clone pair.
\mr{
For fully connected layers used to enhance the ability to detect semantic code clones, we need to fine-tune the code encoder \emph{CC2Vec} and the newly introduced neural networks.
The learning rate and optimizer are the same as above.
Notably, the training dataset utilized is the same dataset used in training the encoder \emph{CC2Vec}, to avoid any leakage of validation and test data into the training process.
}


\subsubsection{Comparative Tools.}
\mr{
We also compare \emph{CC2Vec} with some state-of-the-art code clone detection systems.
The choice of our baseline follows these principles: The method is popular and extensively used, with a high citation in the code clone detection and widely compared in other works. 
}
Specifically, we choose five pretrain-based methods (\ie \emph{Word2Vec} \cite{word2vec}, \emph{Doc2Vec} \cite{doc2vec}, \emph{Code2Vec} \cite{alon2019code2vec}, \emph{CodeBERT} \cite{feng2020codebert}, and \emph{CodeT5} \cite{wang2021codet5}), nine traditional code clone detectors (\ie \emph{SourcererCC} \cite{sajnani2016sourcerercc}, \emph{CCFinder} \cite{kamiya2002ccfinder}, \emph{Nicad} \cite{roy2008nicad}, \emph{Decakrd} \cite{jiang2007deckard}, \emph{CCAligner} \cite{wang2018ccaligner}, \emph{Oreo} \cite{saini2018oreo}, \emph{LVMapper} \cite{wu2020lvmapper}, \emph{NIL} \cite{nakagawa2021nil}, and \emph{Tamer} \cite{hu2023fine}), and six deep-learning-based code clone detectors (\ie \emph{RtvNN} \cite{white2016rtvnn}, \emph{CDLH} \cite{wei2017cdlh}, \emph{TBCNN} \cite{mou2016convolutional}, \emph{ASTNN} \cite{zhang2019astnn}, \emph{SCDetector} \cite{wu2020scdetector}, and \emph{FCCA} \cite{hua2020fcca}).
\mr{
Note that we use the parameters recommended in each paper for these works, and the experimental results are also close to the best result reported in their paper to ensure credibility.
}

\subsubsection{Experimental Environment and Metrics.}

For token categorization, we make use of a Python library (\ie \emph{javalang} \cite{javalang}) to complete our lexical analysis.
After parsing the source code into typed tokens, we leverage \emph{Pytorch} to implement the self-attention mechanism layers and contrastive learning to train the encoder.
Given two vectors of two methods, we use another Python library (\ie \emph{Sklearn} \cite{sklearn}) to compute the cosine similarity.
To measure the effectiveness of \emph{CC2Vec}, we adopt the following widely used metrics. 
Precision is defined as $P = TP/(TP+FP)$.
Recall is defined as $R = TP/(TP+FN)$.
F1 is defined as $F1 = 2*P*R/(P+R)$.
Among them, \emph{true positive} (TP) represents the number of samples correctly classified as clone pairs, \emph{false positive} (FP) represents the number of samples incorrectly classified as clone pairs, and \emph{false negative} (FN) represents the number of samples incorrectly classified as non-clone pairs.
\mr{We employ ten-fold cross-validation to record the F1 score, precision, and recall for each validation run on all experiments.
We calculate the average of these metrics across the ten validations and consider this average as the final performance. 
}

\subsection{\mr{RQ1: Detection Effectiveness}}
\mr{
In this subsection, we evaluate our proposed system by comparing it to three types of code clone detectors, \ie pretrain-based methods, traditional scalable methods, and deep learning-based methods.
Firstly, we assess \emph{CC2Vec}'s ability to detect syntactic code clones compared to other popular pretrain-based methods and traditional scalable tools.
Secondly, we fine-tune \emph{CC2Vec} with neural networks and evaluate its 
semantic code clones detection effectiveness compared to widely used deep learning-based detectors.
}

\subsubsection{\mr{Performance of CC2Vec Compared to Pretrain-based Methods.}}

\mr{
We first analyze our method's effectiveness in detecting syntactic code clones compared to pretrain-based methods.
}
Similar to most of the code clone detectors \cite{sajnani2016sourcerercc}, when we set the detection threshold of \emph{CC2Vec} to 0.7, the precision under this threshold is 98\%.
At the same time, we find that the detection thresholds of different pretrain-based methods are not the same, and to ensure a reasonable comparison, we adapt the thresholds of the other detection methods to make all of them have an accuracy rate of 98\%.
The threshold of \emph{Word2Vec}, \emph{Doc2Vec}, \emph{Code2Vec}, \emph{CodeBERT}, and \emph{CodeT5} is 0.85, 0.75, 0.91, 0.95 and 0.87, respectively.
At the same time, we analyze the recall on cloned code pairs to evaluate the ability of these pretrain-based methods.

\begin{table}[htbp]
\small
\caption{Results of pretrain-based methods and our method \emph{CC2Vec} on BigCloneBench dataset}
\label{tab:big1}
\begin{spacing}{0.8}
\setlength{\tabcolsep}{3mm}{
\begin{tabular}{c|ccccccc}
\toprule
\toprule
\multirow{2}{*}{\textbf{Methods}} & \multicolumn{6}{c}{\textbf{Recall}}   & \multirow{2}{*}{\textbf{Precision}} \\ \cmidrule{2-7}
                      & \textbf{T1}  & \textbf{T2}   & \textbf{VST3}  & \textbf{ST3}  & \textbf{MT3} & \textbf{T4} &                                   \\  
\midrule
Word2Vec   & 1 & 1 & \textbf{0.99} & 0.85 & 0.53  & 0.1   & 0.98  \\ 
Doc2Vec    & 1 & 0.95 & 0.86 & 0.57 & 0.21  & 0.02   & 0.98  \\ 
Code2Vec   & 1 & 1 & 0.95 & 0.82 & 0.54  & 0.17   & 0.98  \\ 
CodeBERT    & 1 & 1 & 0.95 & 0.89 & 0.71  & 0.23   & 0.98  \\ 
CodeT5    & 1 & 1 & 0.98 & 0.91 & 0.77  & 0.49   & 0.98  \\ 
\textbf{CC2Vec}   & \textbf{1} & \textbf{1} & 0.97 & \textbf{0.93} & \textbf{0.81}  & \textbf{0.64}   & \textbf{0.98}  \\ 
\bottomrule
\bottomrule
\end{tabular} }%
\end{spacing}
\end{table}

As shown in Table \ref{tab:big1}, all methods achieved good performance on easier-to-find clones like Type-1, Type-2, and \emph{Very Strongly Type-3} (VST3), being able to detect and recall the vast majority of them.
However, we also see that the recall scores of other pretrain-based methods on \emph{Moderately Type-3} (MT3) and Type-4 perform poorly and are far lower than their precision scores.
Although these code clone detectors model the code language through code corpus, they do not capture the semantic information in the pre-training corpus, but only learn the surface token-based information of the code and obtain an efficient method representation. 
The recall scores of \emph{CodeBERT} and \emph{CodeT5} on Type-4 clones are 23\% and 49\%, respectively, which indicates that they are capable of detecting Type-4 code clones. 
This is reasonable since \emph{CodeBERT} and \emph{CodeT5} are pre-trained on a massive code corpus \cite{feng2020codebert, wang2021codet5, dou2024stepcoder} and have a much higher number of parameters than the other four pretraining-based methods.
\mr{
On the other hand, despite \emph{CodeBERT} and \emph{CodeT5} having a large number of parameters and being pre-trained on extensive code corpora, they do not specifically incorporate the code clone detection task into pre-training tasks, nor do they design optimization goals for code clone detection. 
As a result, they face difficulty in accurately detecting Type-4 code clones.
}


For \emph{CC2Vec}, the experimental results show that the recall score on Type-4 clones is 64\% with 98\% precision, achieving the best performance in pretrain-based methods on BigCloneBench datasets. 
\mr{
This is reasonable that \emph{CC2Vec} considers the potential relationships between tokens, thereby extracting more information than other vanilla pre-trained models during the training phase.
Meanwhile, contrastive learning provides a better way to learn the information between the code snippets and its clone code snippets (\ie positive samples) and other non-clone segments (\ie negative samples) within a batch.
}
Moreover, in contrast to \emph{CodeBERT}, \emph{CC2Vec} do not require a massive pre-training code corpus and a large number of parameters. 
\mr{
\emph{CC2Vec} is only pre-trained on the training dataset, such as \emph{Word2Vec} and \emph{Doc2Vec}.
}

\mr{
In summary, the results indicate that \emph{CC2Vec} outperforms other pretrained-based methods in detecting syntactic code clones, while providing the capability to detect semantic code clones.
However, the ability to detect semantic code clones by only using CC2Vec remains limited.
}

\subsubsection{Performance of CC2Vec Compared to Traditional Tools.}

\mr{
To further examine the effectiveness of \emph{CC2Vec} in detecting syntactic code clones, we also evaluate \emph{CC2Vec} compared to nine traditional scalable detection tools (\ie \emph{SourcererCC} \cite{sajnani2016sourcerercc}, \emph{CCFinder} \cite{kamiya2002ccfinder}, \emph{NiCad} \cite{roy2008nicad}, \emph{Deckard} \cite{jiang2007deckard}, \emph{CCAligner} \cite{wang2018ccaligner}, \emph{Oreo} \cite{saini2018oreo}, \emph{LVMapper} \cite{wu2020lvmapper}, \emph{NIL} \cite{nakagawa2021nil}, and \emph{Tamer} \cite{hu2023fine}).
}

The experimental results are described in Table \ref{tab:big2}.
Many detectors achieve perfect or near-perfect recall for T1 and T2 scenarios, while challenges remain for more semantic clone types like Type-3 and Type-4.
The results show that \emph{SourcererCC} \cite{sajnani2016sourcerercc} achieves very low recall scores on MT3 and Type-4 clones. 
It is reasonable that \emph{SourcererCC} only considers the overlap similarity of tokens between two methods. 
Therefore, it can not handle semantic clones because it does not consider program semantics. Similar performance effects are found in all other traditional code clone detectors (\ie the recall of them on Type-4 clones is approximately zero) that are poor at detecting clone pairs at the semantic level (\ie Type-4 clone pairs).
On the other hand, \emph{CC2Vec} can outperform all other traditional methods on MT-3 and Type-4 clones (\ie the recall is 81\% and 64\%, respectively) and achieve a high precision (\ie the precision is 98\%).


\mr{
In summary, these results further indicate that \emph{CC2Vc} can learn more potential relationship information between tokens to detect syntactic code clones, while enhancing a certain ability to detect semantic code clones.
}

\begin{table}[htbp]
\small
\caption{Results of the other traditional methods and \emph{CC2Vec} on BigCloneBench dataset}
\label{tab:big2}
\begin{spacing}{0.8}
\setlength{\tabcolsep}{3mm}{
\begin{tabular}{c|ccccccc}
\toprule
\toprule
\multirow{2}{*}{\textbf{Methods}} & \multicolumn{6}{c}{\textbf{Recall}}   & \multirow{2}{*}{\textbf{Precision}} \\ \cmidrule{2-7}
                      & \textbf{T1}  & \textbf{T2}   & \textbf{VST3}  & \textbf{ST3}  & \textbf{MT3} & \textbf{T4} &                                   \\  
\midrule
SourcererCC   & 1 & 0.97 & 0.93 & 0.6 & 0.05  & 0   & 0.98  \\ 
CCFinder    & 1 & 0.93 & 0.62 & 0.15 & 0.01  & 0.0   & 0.72  \\ 
NiCad   & 1 & 0.99 & 0.98 & 0.52 & 0.02  & 0   & \textbf{0.99}  \\ 
Deckard    & 0.6 & 0.52 & 0.62 & 0.31 & 0.12  & 0.01   & 0.35  \\ 
CCAligner   & 1 & 1 & \textbf{0.99} & 0.65 & 0.14  & 0   & 0.61  \\ 
Oreo    & 1 & 1 & 1 & 0.89 & 0.3  & 0.01   & 0.89  \\ 
LVMapper   & 1 & 1 & 0.98 & 0.81 & 0.19  & 0   & 0.59  \\ 
NIL    & 1 & 0.97 & 0.88 & 0.66 & 0.19  & 0   & 0.94  \\ 
Tamer & 1 & 1 & 1 & \textbf{0.99} & 0.53  & 0.03   & 0.96  \\ 
\textbf{CC2Vec}   & \textbf{1} & \textbf{1} & 0.97 & 0.93 & \textbf{0.81}  & \textbf{0.64}   & 0.98  \\ 
\bottomrule
\bottomrule
\end{tabular} }%
\end{spacing}
\end{table}


\subsubsection{Performance of CC2Vec Compared to Deep Learning-based Code Clone Detectors.}


\mr{
To further evaluate our proposed method's efficacy in identifying semantic code clones, we carry out comparative experiments on two extensively used datasets (\ie BigCloneBench and GoogleCodeJam), against six popular deep learning-based code clone detection techniques (\ie \emph{RtvNN} \cite{white2016rtvnn}, \emph{CDLH} \cite{wei2017cdlh}, \emph{TBCNN} \cite{mou2016convolutional}, \emph{ASTNN} \cite{zhang2019astnn}, \emph{SCDetector} \cite{wu2020scdetector} and \emph{FCCA} \cite{hua2020fcca}).
Notably, we enhance our method's ability to detect semantic clone pairs by integrating the pre-trained encoder \emph{CC2Vec} with several fully connected layers, as described in Section \ref{sec:clone-detection}. 
The integrated model is subsequently fine-tuned utilizing the identical training dataset employed during the pre-training phase of \emph{CC2Vec}. 
This approach mitigates any potential leakage of validation and test data into the training process, thereby safeguarding the integrity and reliability of the evaluation.
}



\begin{table}[htbp]
\small
\caption{Results of \emph{CC2Vec} and comparative systems for each clone type on BigCloneBench dataset and GoogleCodeJam dataset}
  
\label{tab:big3}
\begin{spacing}{0.8}
\setlength{\tabcolsep}{3mm}{
\begin{tabular}{c|ccc|ccc}
\toprule
\toprule
\multirow{2}[0]{*}{\textbf{Methods}} & \multicolumn{3}{c|}{\textbf{BigCloneBench}} & \multicolumn{3}{c}{\textbf{GoogleCodeJam}} \\ \cmidrule{2-7}
                      & \textbf{Recall}  & \textbf{Precision}   & \textbf{F1}  & \textbf{Recall}  & \textbf{Precision} & \textbf{F1}                                  \\  
\midrule
RtvNN  &   0.01 & 0.95 & 0.01 & 0.90 & 0.20 & 0.33   \\ 
CDLH & 0.74 & 0.92 & 0.82 & 0.70 & 0.46 & 0.55   \\ 
TBCNN   &  0.81 & 0.90 & 0.85 & 0.89 & 0.91 & 0.90  \\ 
ASTNN  &  0.94 & 0.92 & 0.93 & 0.87 & 0.95 & 0.91  \\ 
SCDetector &  0.92 & 0.97 & 0.94 & 0.87 & 0.81 & 0.82 \\ 
FCCA  &  0.92 & \textbf{0.98} & 0.95 & 0.90 & \textbf{0.95} & 0.92  \\ 
\textbf{CC2Vec-1L}   &  0.96 & 0.92 & 0.94 & 0.94 & 0.91 & 0.93   \\ 
\textbf{CC2Vec-3L} &  \textbf{0.97} & 0.94 & \textbf{0.96} & \textbf{0.95} & 0.93 & \textbf{0.94}   \\ 
\textbf{CC2Vec-5L} &  0.97 & 0.94 & 0.96 & 0.95 & 0.93 & 0.94  \\ 
\bottomrule
\bottomrule
\end{tabular} }%
\end{spacing}
\end{table}

\mr{
Table~\ref{tab:big3} describes the Recall, Precision, and F1 scores of our proposed method and six comparative deep learning-based code clone detectors on two datasets.
As in the above experiments, we also use ten-fold cross-validation to record these metrics and report the average scores.
}
\emph{CC2Vec-1L}, \emph{CC2Vec-3L} and \emph{CC2Vec-5L} represent the encoder \emph{CC2Vec} followed by one, three and five fully connected layers, respectively. A ReLU activation layer is inserted in the middle of every two fully connected layers, while for \emph{CC2Vec-1L}, there is no activation layer included. 
\mr{
Among the semantic code clone detectors evaluated,
}
\emph{RtvNN}, \emph{CDLH}, \emph{TBCNN} and \emph{ASTNN} are all AST-based clone detection methods.
As for \emph{RtvNN}, it applies the RNN model to encode both source code tokens and AST to detect code clones. 
However, the tree structure may change significantly even when the code changes slightly. 
The F1 score of \emph{RtvNN} is 1\% and 33\% for BigCloneBench dataset and GoogleCodeJam dataset, respectively.
The other three AST-based methods performed well on both datasets. For example, the F1 of \emph{ASTNN} on the BigCloneBench dataset and GoogleCodeJam dataset achieves 93\% and 91\%, respectively.
\mr{
Another type of semantic code clone detection tools are graph-based, such as \emph{SCDetector} and \emph{FCCA}.
}
These two graph-based methods also have good detection performance. For the BigCloneBench dataset, the recall of \emph{SCDetector} and \emph{FCCA} is 94\% and 82\%, respectively.
As for the GoogleCodeJam dataset, their recall score is 95\% and 92\%, respectively.

For \emph{CC2Vec}, when the fully connected layer number is three (\ie \emph{CC2Vec-3L}), the F1 score is 96\% and 94\% for BigCloneBench dataset and GoogleCodeJam dataset, respectively, achieving comparable performance to other popular deep learning-based methods. 
At the same time, we found that increasing the number of fully connected layers to five (\ie \emph{CC2Vec-5L}) does not result in a significant performance improvement.

\mr{
In summary, the experimental results demonstrate that through fine-tuning the integrated model, our method can effectively detect all types of code clones including syntactic and semantic code clones.
By using token categorization and program encoding, the trainable encoder \emph{CC2Vec} can capture token-level potential relationship information in code segments from code corpus.
Meanwhile, contrastive learning enables the encoder can not only learn information from clone pairs, but also learn from non-clone pairs dynamically constructed during the training phase.
This learning approach enhances the encoder's ability to identify subtle semantic differences between code snippets, to further achieve an improvement in detecting code clones.
}



\subsection{RQ2: Self-Attention and Contrastive Learning}

In this subsection, we conduct three single-factor experiments to validate the use of self-attention mechanism layers and contrastive learning can contribute to the effectiveness of \emph{CC2Vec} on clone detection.
Our first experiment adopts two GRU layers to encode the program source code. 
After training the encoder by using an encoder-decoder model, we can use it to embed other code methods.
The output of the learned encoder is the corresponding vector of a method.
Given two methods, their cosine similarity will be computed as the standard to detect code clones. 
If it is higher than 70\%, they will be reported as a clone pair.
Our second experiment replaces the two GRU layers with two self-attention mechanism layers as the program encoder.
However, we do not perform contrastive learning to train the encoder but use an encoder-decoder model to train it.
Similar to the first experiment, we also compute the cosine similarity of the two methods' vectors to check whether they are a clone pair or not.
Our third experiment is to implement \emph{CC2Vec}.
In other words, we use two self-attention mechanism layers as the program encoder and leverage contrastive learning to train it.
The clone detection phase is the same as in the former two experiments.
In summary, the difference between the first and second experiments is only in the program encoder, and the difference between the second and third experiments is only in the training phase.

\begin{table}[htbp]
\caption{Results on BigCloneBench dataset}
\small
\label{tab:single}
\begin{spacing}{0.8}
\setlength{\tabcolsep}{3mm}{
\begin{tabular}{c|ccc}
\toprule
\toprule
 & \textbf{Recall} & \textbf{Precision} & \textbf{F1} \\  
\midrule
Gated Recurrent Unit (GRU)   & 0.31 & 0.24 & 0.27 \\
Self-Attention   & 0.44 & 0.75 & 0.55 \\
Self-Attention + Contrastive Learning (\emph{CC2Vec})   & 0.74 & 0.98 & 0.84 \\
\bottomrule
\bottomrule
\end{tabular} }%
\end{spacing}
\end{table}

We describe the results of these three comparative experiments in Table \ref{tab:single}.
It shows that the use of self-attention mechanism layers can increase the detection accuracy of \emph{CC2Vec}.
For example, the F1 score of the first experiment is only 27\%, while it increases to 55\% when we adopt self-attention mechanism layers to encode source code.
It is reasonable because the adoption of self-attention mechanism layers can retain some potential relationships between tokens, and these relationships may contain the program details, resulting in better detection effectiveness in detecting code clones.
Moreover, when detecting similar code fragments, self-attention mechanism layers can assign larger weights to similar categories while reducing dissimilar categories' weights.
These weights can boost the ability of code clone detection.

Meanwhile, we also find that integrating contrastive learning enhances the robustness of \emph{CC2Vec} to code changes of different implementations.
Specifically, when we employ contrastive learning in training the program encoder, the F1 score sees a significant jump to 84\%, compared to just 55\% without its use. 
The distance between such semantically similar code pairs is minimized, making it robust to the code changes introduced by different implementations.

In summary, through the results in Table \ref{tab:single}, we see that both self-attention mechanism layers and contrastive learning can boost the ability of \emph{CC2Vec} on code clone detection.

\subsection{RQ3: Interpretability}

\begin{figure}[htbp]
    \centering
    \begin{minipage}{0.46\textwidth}
        \small
        \captionof{table}{Weights of different categories}
        \label{tab:weights}
        \begin{spacing}{0.8}
            \setlength{\tabcolsep}{3mm}
            \begin{tabular}{cc}
                \toprule
                \toprule
                \textbf{Token Category} & \textbf{Attention Weight} \\
                \midrule
                BasicType & 0.1922 \\
                Operator & 0.1739 \\
                Separator & 0.1573 \\
                Identifier & 0.1424 \\
                Modifier & 0.1424 \\
                DecimalInterger & 0.1055 \\
                Keyword & 0.0864 \\
                \bottomrule
                \bottomrule
            \end{tabular}
        \end{spacing}
    \end{minipage}%
    \begin{minipage}{0.26\textwidth}
        \centering
        \includegraphics[width=\textwidth]{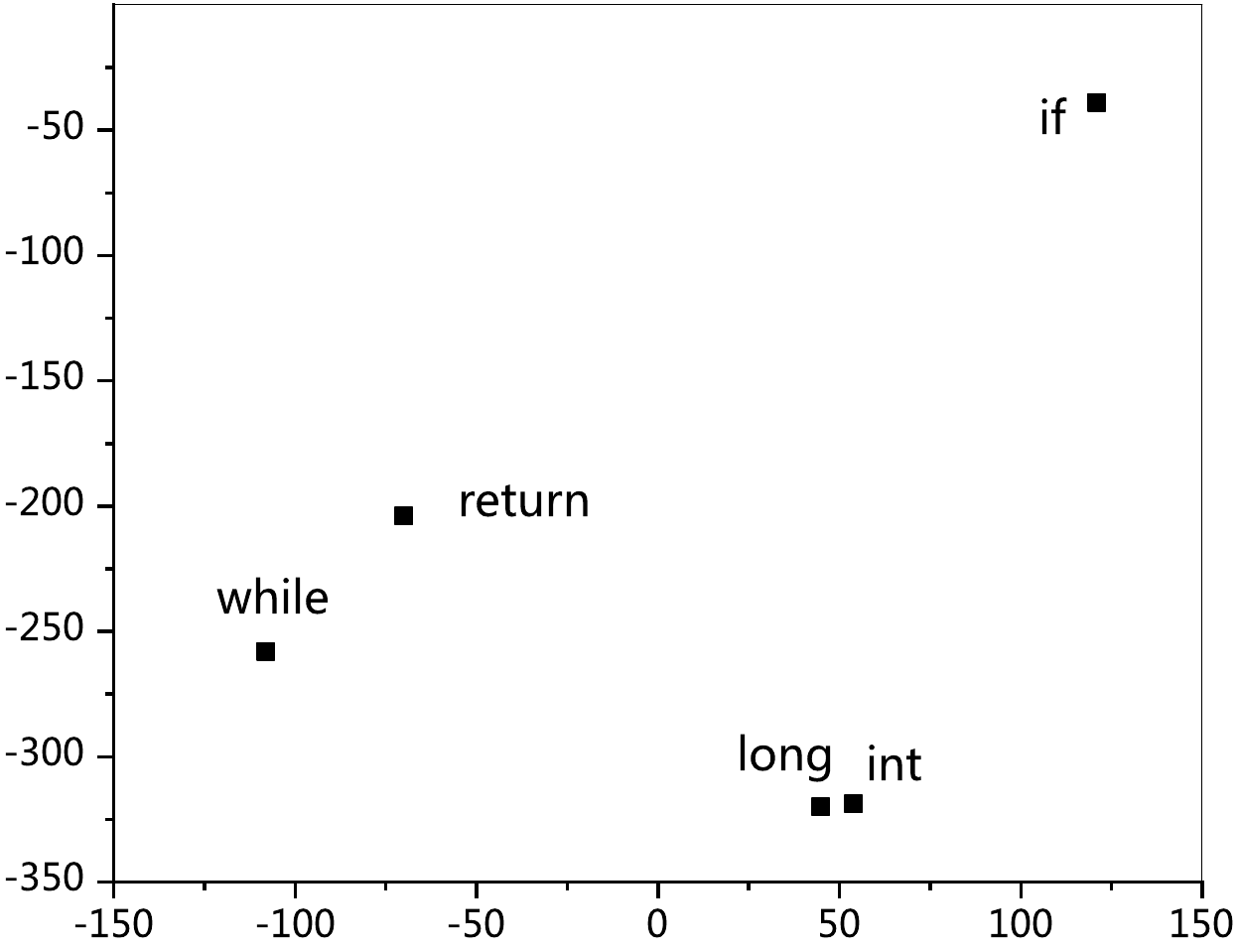}
        \captionof{figure}{The visualization of token vectors}
        \label{fig:visualization}
    \end{minipage}
    \vspace{-1em}
\end{figure}

Since we leverage self-attention mechanism layers as our program encoder, the assigned weights of different categories can interpret the detection results. 
As a matter of fact, the dimension of the attention weight matrix is 15×15, corresponding to 15 categories of tokens, and the value of each position ($i$, $j$) in the matrix represents the correlation between category $i$ and category $j$.
If the matrix column items are summed, and input into a \emph{Softmax()} layer, the importance (\ie weight) of each category can be obtained.
To validate the interpretability of \emph{CC2Vec}, we use methods in List 1 and List 2 as our examples.
Table \ref{tab:weights} shows the weights of different token categories in List 1 calculated from the second self-attention layer.
Since the tokens of the method in List 1 can only be divided into seven categories, the weights of the other eight categories are all 0.

Through the results in Table \ref{tab:weights}, we observe that the weight of \emph{BasicType} token category is the largest, which means that tokens in \emph{BasicType} category have the greatest effect in determining whether List 1 and List 2 are code clones or not.
To further check whether the weights assigned by \emph{CC2Vec} are suitable or not, we choose tokens in \emph{BasicType} category and \emph{Keyword} category as our objects because the weights of these two categories are the largest and smallest in Table \ref{tab:weights}. 
More specifically, we first obtain the corresponding vectors of all tokens in \emph{BasicType} category and \emph{Keyword} category by word2vec \cite{word2vec}.
To show the vectors more intuitively, we perform a visualization technique (\ie TSNE \cite{tsne}) to visualize them in Figure \ref{fig:visualization}.
The figure shows that the vector representations of the token ``int'' and the token ``long'' are similar. 
To obtain more determinate results, we calculate the cosine distance between them.
After calculation, we find that their similarity is 87\%.
For List 1 and List 2, the tokens in \emph{BasicType} category only have token ``int'' and token ``long'', respectively.
Therefore, the similarity between the two categories is the similarity between token ``int'' and token ``long'', which is 87\%.
Such a high similarity makes \emph{CC2Vec} assign a higher weight.
Moreover, we also compute the cosine similarity between the tokens ``while'' and ``if'',
finding a mere 12\% similarity, indicating their dissimilarity.
Meanwhile, results in Figure \ref{fig:visualization} also indicate that the similarity between token ``return'', token ``while'', and token ``if'' are low, resulting in the similarity of \emph{Keyword} category between List 1 and List 2 is low.
It is the reason why \emph{CC2Vec} assigns a smaller weight to \emph{Keyword} category than \emph{BasicType} category.

In summary, incorporating self-attention mechanism layers in CC2Vec provides interpretability for code clone detection outcomes.

\subsection{RQ4: Scalability}

Our final experiment aims to examine the ability of \emph{CC2Vec} on large-scale code analysis.
Specifically, we randomly choose one million code pairs from BigCloneBench and run \emph{CC2Vec} and our comparative systems on them to collect the runtime overheads.
We run these tools three times and report the average as their final runtime.
For pretrain-based methods (\ie \emph{Word2Vec}, \emph{Doc2Vec}, \emph{Code2Vec}, \emph{CodeBERT}, \emph{CodeT5}, and \emph{CC2Vec}), the training time indicates the time for pre-trained with the code corpus and the prediction time is the sum of the time to obtain vectors and the time spent in computing the cosine similarity.
For example, we should perform contrastive learning to train a self-attention-based program encoder first. Otherwise, we cannot detect code clones by using \emph{CC2Vec}.
For deep learning-based methods (\ie \emph{RtvNN}, \emph{CDLH}, \emph{TBCNN}, \emph{ASTNN}, \emph{SCDetector}, \emph{FCCA} and \emph{CC2Vec-3L}), they need to train a model first, and then the model can be used to detect code clones.
In other words, their runtime overheads are divided into two parts, corresponding to training time and prediction time.

\begin{table}[htbp]
\begin{spacing}{0.2}
\small
  \centering
  \caption{Time performance of code clone detectors on analyzing one million code pairs}
    \begin{tabular}{c|c|cc}
    \toprule
    \toprule
    \multicolumn{2}{c|}{Tools} & Prediction Time & Training Time \\
    \midrule
    \multirow{4}[8]{*}{Pretrain-based methods} & \multicolumn{1}{c|}{Word2Vec} &   25s    & 7,523s \\
\cmidrule{2-4}          & \multicolumn{1}{c|}{Doc2Vec} &   28s    & 9,126s \\
\cmidrule{2-4}          & \multicolumn{1}{c|}{Code2Vec} &   30s    & 17,345s \\
\cmidrule{2-4}          & \multicolumn{1}{c|}{CodeBERT} &   274s    & - \\
\cmidrule{2-4}          & \multicolumn{1}{c|}{CodeT5} &   873s    & - \\
    \midrule
    \multirow{9}[18]{*}{Traditional methods} & SourcererCC & 16s   & 0s \\
\cmidrule{2-4}          & CCFinder & 23s   & 0s \\
\cmidrule{2-4}          & NiCad & 14s   & 0s \\
\cmidrule{2-4}          & Deckard & 72s   & 0s \\
\cmidrule{2-4}          & CCAligner & 10s   & 0s \\
\cmidrule{2-4}          & Oreo  & 15s   & 0s \\
\cmidrule{2-4}          & LVMapper & 11s   & 0s \\
\cmidrule{2-4}          & NIL   & 9s    & 0s \\
\cmidrule{2-4}          & Tamer & 10s   & 0s \\
    \midrule
    \multirow{6}[12]{*}{Deep learning-based methods} & RtvNN & 35s   & 5,206s \\
\cmidrule{2-4}          & CDLH  & 90s   & 45,317s \\
\cmidrule{2-4}          & TBCNN & 86s   & 41,168s \\
\cmidrule{2-4}          & ASTNN & 2,894s & 16,096s \\
\cmidrule{2-4}          & SCDetector & 139s  & 2,937s \\
\cmidrule{2-4}          & FCCA  & 91s   & 56,769s \\
    \midrule
    \multirow{2}[4]{*}{Ours} & CC2Vec &   28s    & 3,649s \\
\cmidrule{2-4}          & CC2Vec-3L &   83s    & 4,871s \\
    \bottomrule
    \bottomrule
    \end{tabular}%
  \label{tab:overhead}%
  \end{spacing}
  \vspace{1em}
\end{table}%

The runtime overheads of our proposed methods (\ie \emph{CC2Vec} and \emph{CC2Vec-3L}) and other detection systems are presented in Table \ref{tab:overhead}.
\emph{CC2Vec} is trained faster than other pretrain-based methods.
For example, the training time of \emph{Code2Vec} is about 17,345 seconds, and \emph{CC2Vec} takes about 3,649 seconds.
The training process for \emph{CodeBERT} is notably resource-intensive and time-consuming \cite{feng2020codebert, wang2021codet5}, hence its training duration is not measured.
\mr{
The prediction time of \emph{CodeT5-base} exceeds that of \emph{CodeBERT}.
It is reasonable that \emph{CodeT5} has a greater number of parameters compared to \emph{CodeBERT} and it is also an encoder-decoder model \cite{wang2021codet5}.
}
Since traditional methods do not need to train, their training time is zero.
As for prediction time, the token-based method \emph{NIL} consumes the least runtime because it only uses an N-gram representation of token sequences for clone detection.
Compared to \emph{NIL}, \emph{CC2Vec} and \emph{CC2Vec-3L} take more runtime to predict the same code pairs.
It is reasonable because \emph{CC2Vec} conducts lexical analysis to divide the tokens into different categories, which consumes some time to complete the step. And \emph{CC2Vec-3L} fine-tuned based on \emph{CC2Vec} combined with neural networks.
Compared to deep learning-based methods, \emph{CC2Vec-3L} are more scalable.
It takes about 4,871 seconds to finish the training procedure and 83 seconds to detect one million code pairs.
The training phase consumes more time than \emph{SCDetector} (\ie 2,937 seconds) and more prediction time than \emph{RtvNN} (\ie 35 seconds), but is faster than other deep learning-based cloned code detection tools.
Compared to \emph{RtvNN}, whether from the training or prediction phase, the runtime overhead of \emph{CC2Vec} is lower.
For \emph{ASTNN}, it requires 2,894 seconds to predict the one million code pairs, which is about 34 times slower than \emph{CC2Vec-3L}.
\mr{
As for \emph{FCCA}, our method has a significant advantage in detection efficiency during the training phase. 
In addition, the detecting time of \emph{CC2Vec-3L} is also less than \emph{FCCA} (\ie the detecting time of \emph{CC2Vec-3L} and \emph{FCCA} are 83s and 91s, respectively).
}

In summary, due to self-attention mechanism layers and contrastive learning, \emph{CC2Vec} and \emph{CC2Vec-3L} are not as fast as traditional methods.
\mr{
However, it consumes less time than most other pretrain-based methods and deep learning-based methods and also can achieve comparable performance to these deep learning-based methods.
Efficient detectors can reduce time and resource consumption and have the potential to detect large-scale code clones.
}

\section{Discussions}

\subsection{Threats to Validity}
\mr{
The first threat comes from the BigCloneBench dataset, a recent study \cite{krinke2022bigclonebench} has shown that the dataset has some quality issues which may cause negative effects on \emph{CC2Vec}.
To mitigate the threat and make our experimental results more trustworthy, we also validate \emph{CC2Vec} on another widely used code clone benchmark (\ie the Google Code Jam dataset). The experimental results still show that \emph{CC2Vec} can perform well on detecting code clones.
The second threat comes from the Google Code Jam dataset, it may contain several inaccuracies since some participants may have misunderstood the task and implemented something semantically different. 
To mitigate the threat, we randomly choose 10\% samples from each competition problem and manually verify the functionality. 
The results show that these samples all correctly address the problem. 
The third threat comes from the token types, in the token categorization phase, we select a total of 15 categories to commence our experiments. 
The selection of these categories may cause some inaccuracies since the total number of token categories parsed by \emph{javalang} is not clear. 
To mitigate the situation, we perform a statistical analysis to select the categories with a high number of occurrences and add a \emph{Null} category to represent the remaining categories.
The fourth threat comes from the runtime overhead. The calculation of runtime overheads of \emph{CC2Vec} and its comparative tools may also cause some inaccuracies due to the different machine states such as CPU usage. 
We mitigate the threat by conducting evaluations three times and reporting the average runtime overhead in our paper.
}


\subsection{Differences From C4}
\mr{
The most similar work to \emph{CC2Vec} is \emph{C4} \cite{tao2022c4} which uses contrastive learning to detect cross-language code clones.
However, our method greatly differs from \emph{C4}. 
Firstly, \emph{CC2Vec} processes code snippets through syntactic analysis, tokenization, and categorization. 
Secondly, it utilizes a two-layer attention mechanism to capture the relationships between tokens of the same category and different categories within the language.
Finally, we use contrastive learning to derive the code vector. 
Our primary contribution is how we capture the relationships among tokens in the code, while the subsequent contrastive learning aims to obtain a better code vector.
In contrast, \emph{C4} directly tokenizes code snippets and uses CodeBERT for representation, with training via contrastive learning. 
Contrastive learning is a general method, but our contribution is to capture the relationships between tokens, which is not explored in \emph{C4}. 
Additionally, our research field differs from that of \emph{C4}. 
\emph{C4} investigates cross-language clone detection, whereas \emph{CC2Vec} focuses on obtaining better code vectors for more rapid clone detection. 
We do not use CodeBERT, a large model, but instead train with a lightweight network having only two layers of attention, making our detection more efficient. 
}

\subsection{Why does CC2Vec Perform Better?}
The reasons why \emph{CC2Vec} is superior to other token-based systems are mainly two-fold.
First, \emph{CC2Vec} uses self-attention mechanism layers as the program encoder which can not only compute the corresponding weights to interpret the detection results but also retain some potential relationships between tokens.
\mr{
It is just like a data flow, when some tokens are defined or used earlier, the attention mechanism can establish connections between the definition and the reference locations. 
These connections can better aid in the detection of clones. 
In addition, the attention mechanism is a double-direction neural network, which means that it can capture the relationships between tokens better.
This method is similar to language in natural language processing, where attention can establish connections between two tokens that are far apart.
These potential relationships between tokens can retain some program details between source code tokens.
}
Second, the use of contrastive learning in \emph{CC2Vec} can resist the changes of code structures introduced by different implementations, making it possible to detect some semantic code clones.



\section{Related Work}

According to the extraction of different representations, existing code clone detection methods can be divided into five main categories: text-based, token-based, tree-based, graph-based, and metrics-based approaches.

Text-based methods \cite{dou2023towards, ducasse1999language, roy2008nicad, wettel2005archeology, lee2005sdd, barbour2010technique, marcus2001identification} compute the similarity between two code fragments in the form of text.
Wettel et al. \cite{wettel2005archeology} propose a method for clone detection based on code line comparison. 
Lee et al. \cite{lee2005sdd} design \emph{SDD} to detect large-scale code clones. 
Token-based methods \cite{kamiya2002ccfinder, gode2009incremental, sajnani2016sourcerercc, li2017cclearner, wang2018ccaligner, murakami2013gapped} need to transform the source code into tokens by lexical analysis, then the similarity can be computed by analyzing tokens.
Murakami et al. \cite{murakami2013gapped} use a hash algorithm to convert the code fragment into a token sequence and introduce a classic and highly efficient Smith-Waterman algorithm to detect code clones.
Sainj et al. \cite{sajnani2016sourcerercc} implement \emph{SourcererCC} to capture the tokens' overlap similarity between different methods to detect near-miss Type-3 clones. 
\emph{SourcererCC} is the most scalable code clone detector which can scale to scan more than 428 million files on Github \cite{lopes2017dejavu}.
Due to the lack of program details, these text-based and token-based methods are difficult to handle semantic clones.

Tree-based methods \cite{jiang2007deckard, wei2017cdlh, zhang2019astnn, baxter1998clone, yang1991identifying, wahler2004clone, koschke2006clone, wu2022detecting, hu2022treecen} need to extract the \emph{abstract syntax tree} (AST) of a code method and then use it to detect code clones.
Baxter et al. \cite{baxter1998clone} design a tree-based code clone detector \emph{CloneDR}.
It first uses a compiler tool to generate the AST of the code, then calculates the hash code of each subtree, and finally performs similarity matching based on the obtained hash code.
Wei et al. \cite{wei2017cdlh} introduce \emph{CDLH} which leverages Tree-LSTM \cite{tai2015treelstm} on generated binary trees to encode them.
Although tree-based methods can detect semantic code clones, however, they suffer from low scalability because of the large execution times \cite{zhang2019astnn}.
Graph-based methods \cite{krinke2001duplix, komondoor2001pdgdup, wang2017ccsharp, zhao2018deepsim, chen2014centroid, wu2020scdetector, shan2023gitor} distill the program semantics into different graph representations, such as \emph{control flow graph} (CFG) and \emph{program dependency graph} (PDG).
Komondoor et al. \cite{komondoor2001pdgdup} and Krinke et al. \cite{krinke2001duplix} both use PDG as the standard to detect code clones.
They apply graph analysis to discover isomorphic subgraphs and use these subgraphs to detect semantic code clones.
Wu et al. \cite{wu2020scdetector} apply centrality analysis to transform the CFG of a method into certain semantic tokens and detect code clones by analyzing these semantic tokens.
Graph-based methods can also detect semantic code clones, but due to the complexity of graph isomorphism and the heavy-weight time consumption of graph matching, they can not scan codes on a large scale.


\section{Conclusion}

\mr{
In this paper, we propose to boost the capability of code clone systems to efficiently detect syntactic and semantic code clones.
To complete the purpose, we introduce \emph{CC2Vec}, a code encoding method for code clone detection.
\emph{CC2Vec} uses two self-attention mechanism layers as the program encoder and applies contrastive learning to train the encoder.
We can detect simple code clones by directly utilizing \emph{CC2Vec}, while identifying more complicated code clones by combining it with a few neural networks.
}
Through the experimental results, we find that \emph{CC2Vec} can detect more code clones than five pretrain-based methods and 15 state-of-the-art code clone detectors.
As for scalability, although \emph{CC2Vec} is not as fast as traditional methods, it surpasses the majority of pretrain-based and deep learning-based methods.



\bibliographystyle{ACM-Reference-Format}
\bibliography{Toka}

\end{document}